\documentclass[twocolumn]{webofc}

\usepackage[varg]{txfonts}   
\usepackage{xcolor}
\begin{document}

\title{
\vspace*{-3cm} Grain size effect on the compression and relaxation of a granular column:  solid particles vs dust agglomerates}
%
%

\author{\firstname{Felipe} \lastname{Pacheco-V\'azquez}\inst{1,4}\fnsep\thanks{\email{fpacheco@ifuap.buap.mx}} \and
        \firstname{Tomomi} \lastname{Omura}\inst{2,4}\fnsep\thanks{\email{tomura@edo.osaka-sandai.ac.jp}} \and
        \firstname{Hiroaki} \lastname{Katsuragi}\inst{3,4}\fnsep\thanks{\email{katsuragi@ess.sci.osaka-u.ac.jp}}
}

\institute{Instituto de F\'isica, Benem\'erita Universidad Aut\'onoma de Puebla, Apartado Postal J-48, Puebla 72570, Mexico
\and
           Institute of Education Center of Advanced Education, Osaka Sangyo University, 3-1-1 Nakagaito, Daito-shi, Osaka 574-8530, Japan
\and
           Department of Earth and Space Science, Osaka University, 1-1 Machikaneyama, Toyonaka 560-0043, Japan
\and
Department of Earth and Environmental Sciences, Nagoya University, Furocho, Chikusa, Nagoya 464-8601, Japan
          }

\abstract{We studied experimentally the effect of grain size and maximum load on the compaction and subsequent relaxation of a granular column when it is subjected to vertical uniaxial compression. The experiments were performed using two different types of grains: 1) solid glass beads, and 2) porous beads that consist of agglomerates of glass powder. We found that the compression force increases non-linearly with time, with sudden drops for the case of glass beads and periodic undulations for dust particles. Whereas the grain size effect is small in the average force load, the fluctuations become larger as the grain size increases. On the other hand, the relaxation process is well described by the Maxwell model with three different relaxation time scales. 
}

\maketitle
\section{Introduction}
\label{intro}
The relaxation process of a granular material after being continuously compressed is not totally well understood. This is reflected for instance in the undesirable sinking of airports  and highways \cite{Mola:2018}, where the relaxation times result considerably shorter than those estimated from existing models (e.g. the Maxwell model). This is in part because the relaxation dynamics strongly depends on the compression process and on the particles nature (soft, rigid, hierarchical, cohesive), as it has been revealed by different studies focused on the uniaxial vertical compression of granular materials and powders \cite{Brujic:2005,Valdes:2012,Barraclough:2016,Guillard:2015, Wunsch2019, Katsuragi2020}. 

Acccording to Ref. \cite{Brujic:2005}, the stress relaxation in a granular column of rigid glass beads previously subjected to slow oscillatory compaction decreases logarithmically; however, at large compaction rates, a fast exponential decay followed by logarithmic relaxation was observed \cite{Brujic:2005}. For the case of brittle materials, propagation of upward compaction bands was reported during vertical compression of cereals \cite{Valdes:2012} and snow \cite{Barraclough:2016}. Numerical simulations were able to reproduce such propagation bands and classify them in different compaction regimes in terms of two non-dimensional groups: the elasto-breakage number and visco-breakage number  \cite{Guillard:2015}. These propagating bands are characterized by sudden variations in the compression stress produced by the breakage of particles at the bottom of the bed, that allows the relaxation of the particles and their further rearrangement. In contrast, the compression of columns composed of flexible rubber grains was characterized by a smooth rather than periodic deformation process with a pronounced compaction near the compressor (a piston) that gradually diffuses with depth \cite{Valdes:2012}.

For the case of hierarchical granular matter \cite{Katsuragi2020}, we reported recently that the compression of granular columns composed of 1 mm dust aggregates was characterized by periodic undulations of the force load with wavelengths dependent on the compression rate, followed by a multistage relaxation with different time scales. This dynamics emerges from the combination of brittle and elastic features of the dust aggregates, and it was in clear contrast with sudden force drops that characterize the compression process of solid glass beads columns.  

In this paper, we extend our previous research \cite{Katsuragi2020} by exploring the effect of the grain size on the compression and relaxation dynamics of a granular column composed of solid beads or hierarchical granular matter. We found that the increase of the average compression force is not very sensitive to the particle size. In contrast, the amplitudes of the sudden force drops for glass beads and undulations for dust particles increase notably with the grain size. For both materials, the relaxation dynamics is well described by the Maxwell model using three exponential terms with short, medium and long characteristic times.

\section{ Materials and experimental procedure}
\label{sec-1}

For experiments with glass beads (density $\rho=2.5$ g/cc), we used particles (supplied by AS-ONE Corp) of five different representative sizes $D_g$: 0.1 mm (0.105-0.125 mm), 0.4 mm (0.350-0.500 mm), 0.6 mm (0.500-0.710 mm), 0.8 mm (0.710-0.990 mm) and  2 mm (1.500-2.500 mm). For experiments with dust particles, we first produced the dust agglomerates as follows:  powder composed of tiny glass beads of 5 $\mu$m in diameter was poured into a rotating drum and sprayed with water, the rotation process induced the formation of quasi-spherical powder aggregates of different sizes due to the cohesion induced by liquid bridges. The resulting particles were left to dry at ambient temperature (after that, the aggregates have strength keeping their shapes because molecular water always exists as long as we perform the experiments under the atmospheric pressure environment, and at the same time, van der Waals forces can contribute to the cohesion in micrometric scale). Finally, the agglomerates were carefully sieved to separate them by size in four ranges of diameters: $D_d <0.6$ mm, 0.6-1.4 mm, 1.4-2 mm and 2-4 mm. The porous dust particles resulting from the above process had in average a grain packing fraction $\phi_{d}^{grain}\approx 0.30$.

\begin{figure}[h]
\centering
\includegraphics[width=7.5cm,clip]{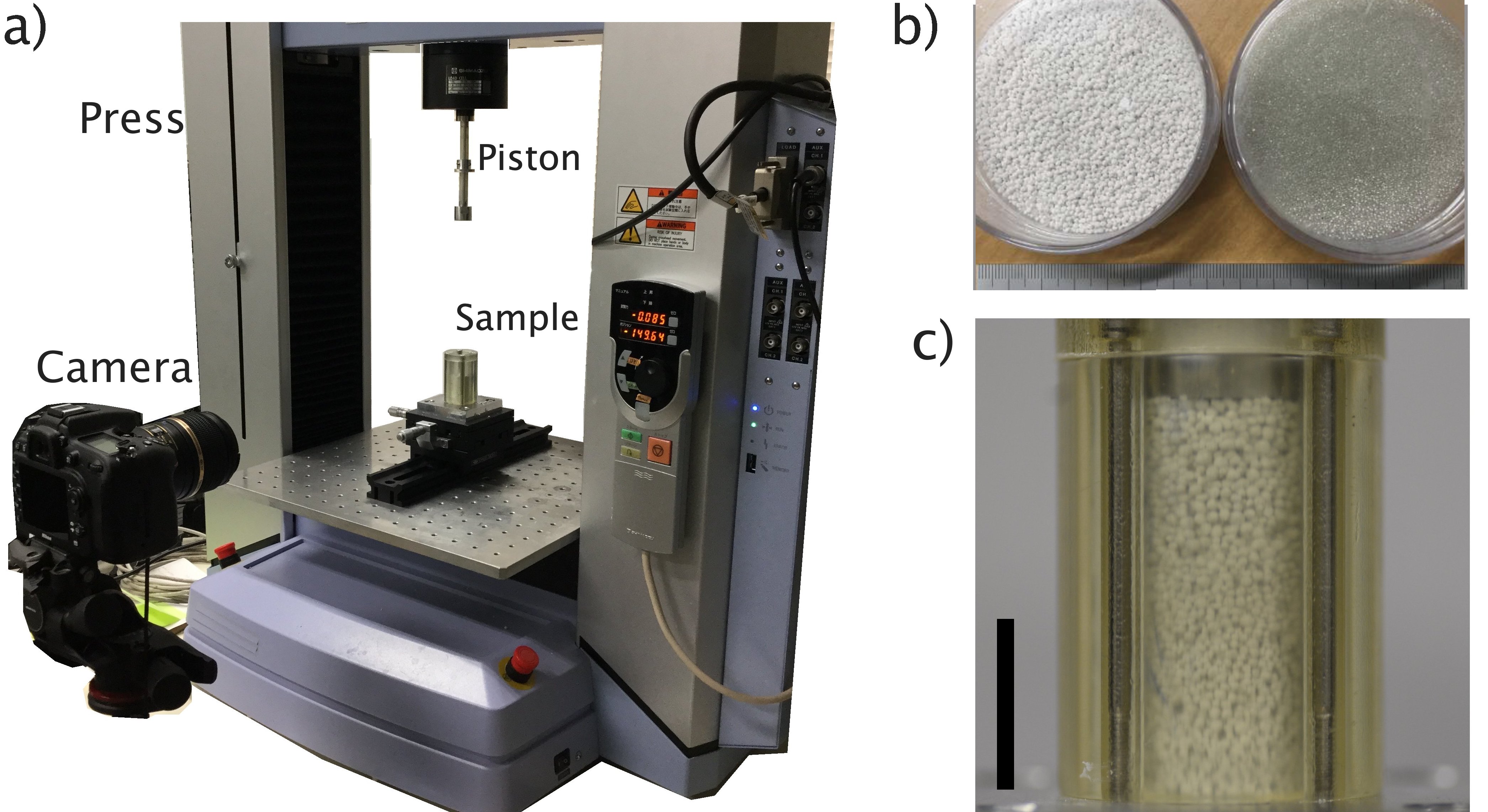}
\caption{a) Experimental setup used to measure the force load $F$ vs piston stroke $z$.  b) Dust particles and glass beads used in the experiments. In this study, the parameters and results associated with glass beads and dust aggregates are labeled with subscripts \textit{g} and \textit{d}, respectively. c) Column of dust particles inside the cylindrical container previous compression (scale bar: 20 mm). }
\label{fig-1}   
\end{figure}
\vspace*{-0.4cm}
Figure \ref{fig-1} shows the experimental setup used in our experiment. A mass $m_g=15~g$ of glass beads or $m_d=8~g$ of dust aggregates was poured into a transparent acrylic cylindrical vessel of 20 mm inner diameter (10 mm wall thickness) and 57 mm high, obtaining similar bed packing fractions, $\phi_{bed} \sim 0.59$, for both materials. However, since dust particles are porous, the bulk packing fraction of the bed was $\phi_{d}^{bulk}=\phi_{bed} \phi_{d}^{grain} \sim 0.18$. The sample was compressed in the direction of gravity with an aluminum piston using a Shimadzu AG-X 100N  uniaxial press. In all the tests, the compression velocity $v$ was kept constant during the process, with magnitude $v= 0.1$ mm/min. Initially, the piston stroke $z$ was set to zero when the piston touched the upper surface of the sample, then the sample was compressed until reaching a maximum compression force $F_{max}$ previously selected (which values were 1, 5, 10, 20, 30 and 100 N). Data was registered at a frequency of 100 Hz obtaining  the load force $F$ as a function of time $t$, namely, $F(t),$ or equivalently $F(z)$ (where $z=v t$). 
When $F_{max}$ was reached, the piston was stopped and kept fixed while $F(t)$ was registered during at least one hour of the relaxation process. Since the dust particles were deformed or broken during the compression, they were replaced after each test.  Although the maximum applied stress ($\sim 300$ kPa for $F_{max}=100$ N) was well below the typical glass strength ($\sim ~ 7$ MPa),  the glass beads were also replaced after each test to avoid attrition. The room temperature, $T=25^\circ$C, and humidity,  $\sim 50\%$, were controlled and kept constant during each experiment.

\section{Results}
\label{sec-2}

\subsection{Glass beads}
\label{sec2-1}
\begin{figure}[ht!]
\includegraphics[width=8cm]{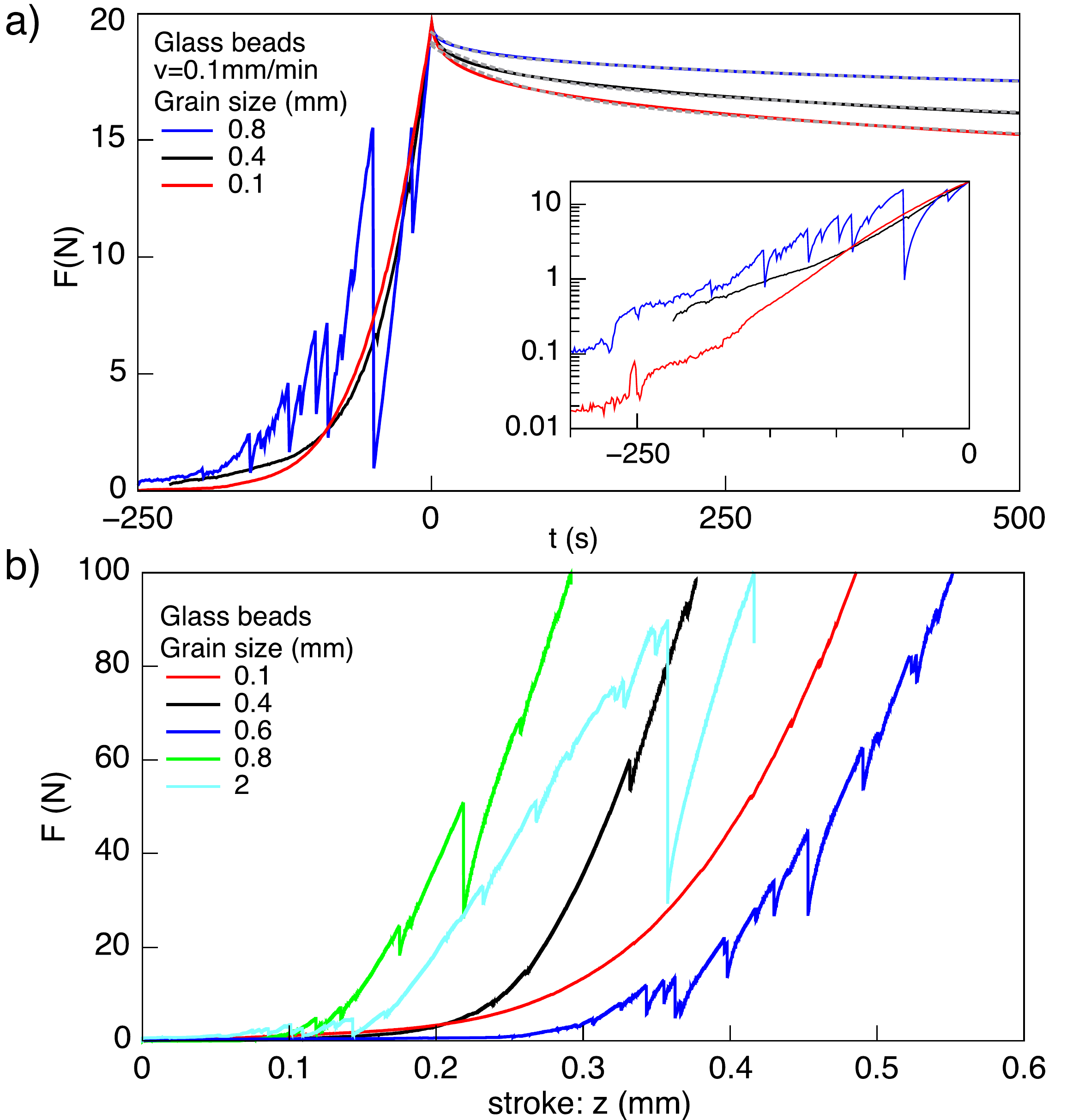}
\caption{ a) $F$ vs $t$ for three values of grain size: $t<0$ corresponds to the compression phase, $t=0$ s indicates when the maximum compression load $F_{max}=20$ N is attained, and $t>0$ corresponds to the relaxation phase.  Inset: corresponding log-linear plot of $F$ vs $t$ during compression suggesting an exponential growth of $F(t)$. b) $F$ vs $z$ during compression up to $F_{max}=100$ N for different values of grain size. 
}
\label{fig-2}
\end{figure}

Figure \ref{fig-2}a  shows the force load $F$ vs $t$ obtained during the compression of granular columns of glass beads of three different sizes followed by the relaxation process.  Negative values of $t$ corresponds to the compression, and  $t=0$ s marks the onset of the relaxation from the maximum load $F_{max}= 20$ N. For grains of diameter $D_g=0.1~mm$, the compression curve is smooth and continuous (red curve), in contrast to the abrupt force drops that appear intermittently for larger grain sizes. Note that the magnitude of these drops becomes more important as the grain size increases. On the other hand, the relaxation is characterized by a continuous decrease of $F(t)$ in all cases, and it is faster for smaller grains. In Fig. \ref{fig-2}b, the compression curves are shown for a larger value of $F_{max}=100$ N that allows to better visualize the difference in the abrupt force drops for different grain size in a wider range. Note again that the force fluctuations are negligible for grains of $0.1$ mm but become more and more pronounced for larger grains. 


\begin{figure*}[ht!]
\centering
\includegraphics[width=16cm]{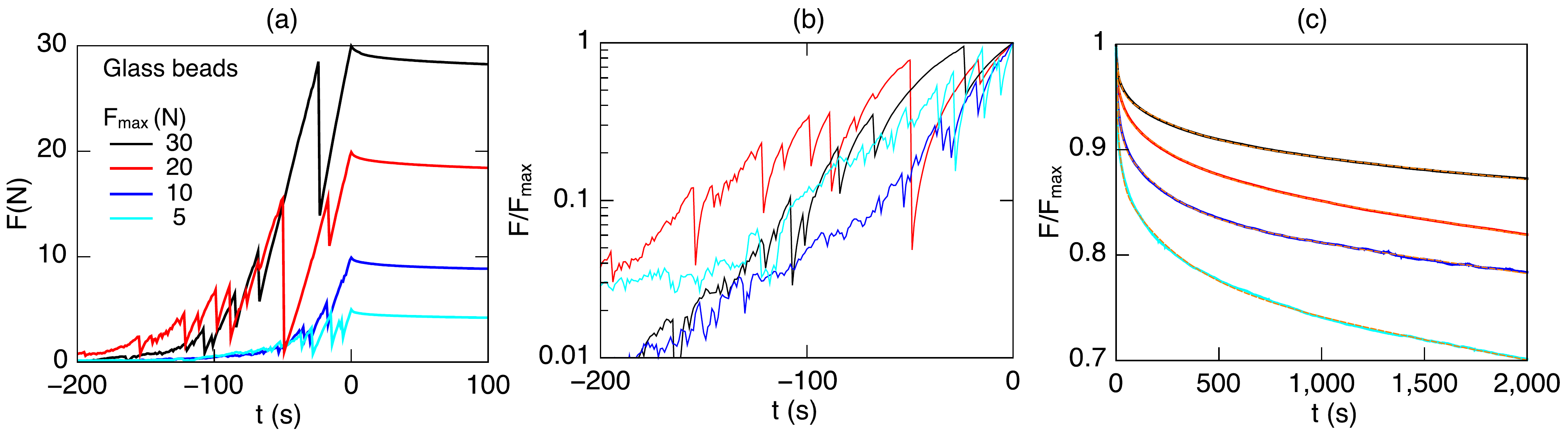}
\vspace*{-0.4cm}
\caption{\textit{Glass beads:} a) $F$ vs $t$ for beads of $D_g=0.8$ mm compressed until different values of $F_{max}$. b) Corresponding Log-linear plot of $F$ vs $t$ during the compression process ($t<0$). c)  $F$ normalized by $F_{max}$ vs time during the relaxation process ($t>0$). Solid lines correspond to the experimental measurements and dashed orange lines represent the best fit of the experimental data using Eq. \eqref{Eq1}.
}
\label{fig-3}
\end{figure*}

It can also be noticed in Fig. \ref{fig-2}b that the curves are offset from each other. Since the initial stroke $z=0$ was selected when the piston touches the grains at the top of the bed (producing an increase in $F$ from the noise of the signal), and considering that glass beads are rigid and cannot be deformed with the applied stress, the subtle structure of the initial configuration of the grains affects the behavior of early stage compression. Therefore, the variation of the initial grains-network structure results in the dispersion of the compression force in the early stage. Note also that $F_{max}= 100$ N is reached in less than 0.5 mm of vertical compression.

Figure \ref{fig-3}a shows $F$ vs $t$ for granular columns of glass beads of $D_g=0.8$ mm compressed until reaching different values of $F_{max}$. Again, the measured force increases with abrupt drops  that become more notorious as the bed is compressed. It must be noticed that the fluctuations of $F$ are independent of the maximum force load but they become larger as the mean value of $F(t)$ increases. On the other hand,  at $t=0$ s, the bed relaxation begins from $F=F_{max}$ and it is characterized by a monotonous continuous decrease of $F(t)$. In this case, the relaxation seems faster for smaller values of $F_{max}$, as it is shown in Fig. \ref{fig-3}c.

\subsection{Dust particles}
\label{sec-3}
\begin{figure}[h]
\centering
\includegraphics[width=7.3cm,clip]{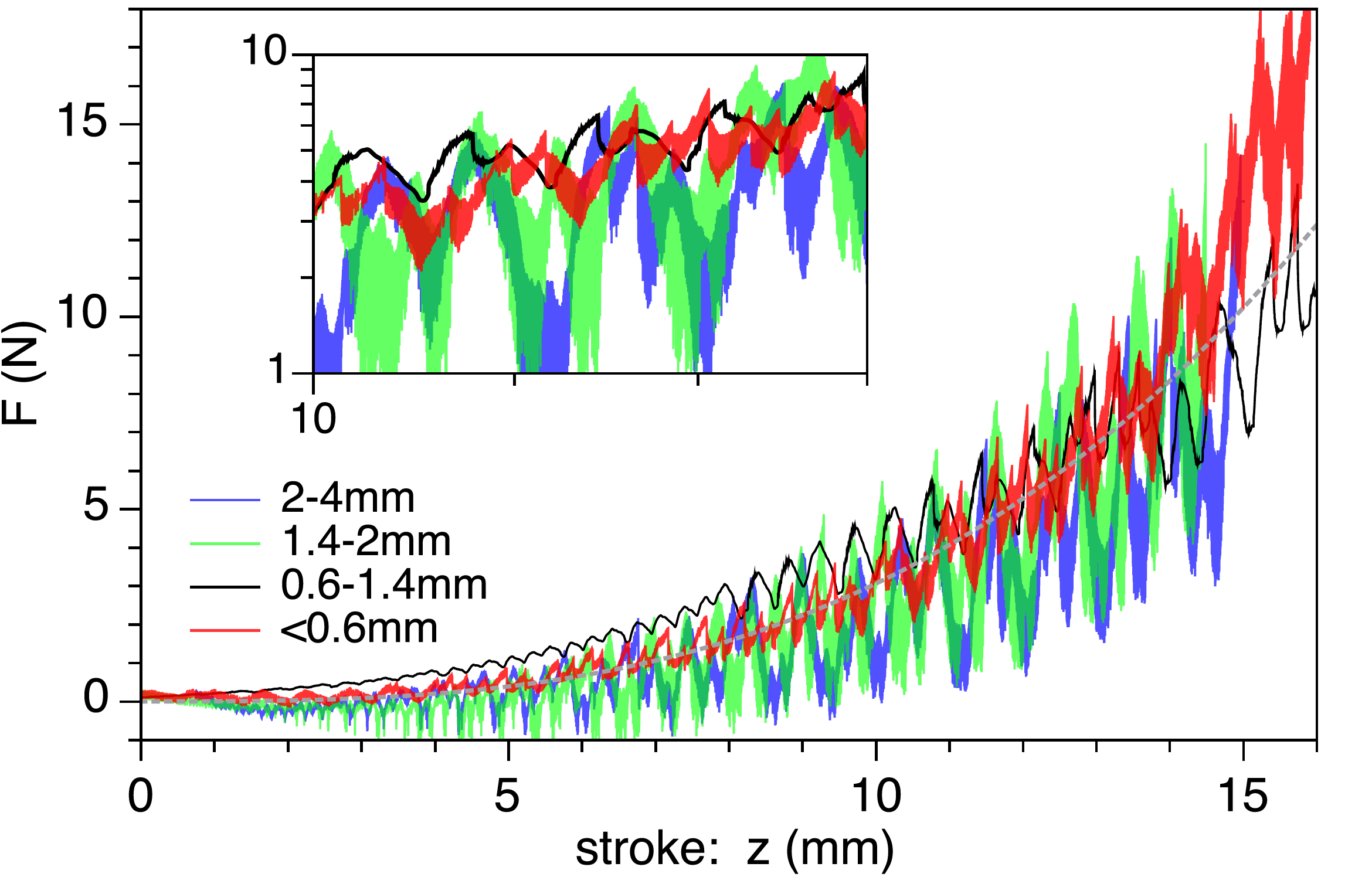}
\vspace*{-0.3cm}
\caption{\textit{Dust particles:} Force load $F$  vs piston stroke $z$ for different particle sizes. The dashed gray line indicates a power-law fit of the form $F(z)=A z^p$, where $A=0.003$ N/m$^3$ and $p=3$. Inset: Zoom of the log-log plot of $F$ vs $z$ that helps to visualize the increase of the undulation amplitude with the grain size.  }
\label{fig-4}   
\end{figure}

The results of $F$ vs $z$ for the compression of dust particles  until reaching a maximum force load of 20 N are shown in Fig. \ref{fig-4}. In this case, the piston stroke required to reach $F_{max}$ is $z \sim 16$ mm, which contrasts to $z < 1$ mm required for the case of solid glass beads. This reveals that the particles are considerably deformed and broken during the compression process.  Moreover, oscillatory fluctuations of $F(z)$ were measured instead of abrupt drops. These results  are in agreement with our previous findings for different compression velocities of dust particles of a given size reported in \cite{Katsuragi2020}. The amplitude of the oscillations in $F(z)$ increases with the grain size and seems proportional to the mean value of $F(z)$. The wavelength of the oscillations also increases with the grain size (see inset in Fig. \ref{fig-4}). On the other hand, compared to the case of glass beads, the stress relaxation for dust particles (not shown) is markedly faster during the first seconds,  but then, it decreases also smoothly and continuously, as it was reported for a fixed grain size in Ref. \cite{Katsuragi2020}.

\section{Analysis and discussion}
\label{sec-3}
\subsection{Compression process} \label{sec3-1} 
For the case of glass beads, figs. \ref{fig-2}a(inset) and \ref{fig-3}b show log-linear plots of $F$ vs $t$ for different values of $D_g$ and $F_{max}$, respectively. Since the piston stroke is $z=vt$, the observed linear behavior suggests an exponential growth of the form $F \propto \exp(z/z_g)$, where $z_g$ is a fitting parameter, in agreement with previous results found using different compression rates. This dependence is discussed in detail in Ref. \cite{Katsuragi2020}. The compression force is independent of $F_{max}$ since one curve with a large value of $F_{max}$ would correspond to the continuation of another curve with smaller value of $F_{max}$. Accordingly, all the curves have approximately the same slope in Fig. \ref{fig-3}b. 
The grain size has no important effect on the mean value of $F$ in the studied range, as it can be noticed in Fig. \ref{fig-2}a where the three curve are superimposed. Since all the glass beads are made of the same material they have similar stiffness and the resistance to compression is approximately the same.


\begin{figure}[ht!]
\includegraphics[width=8cm]{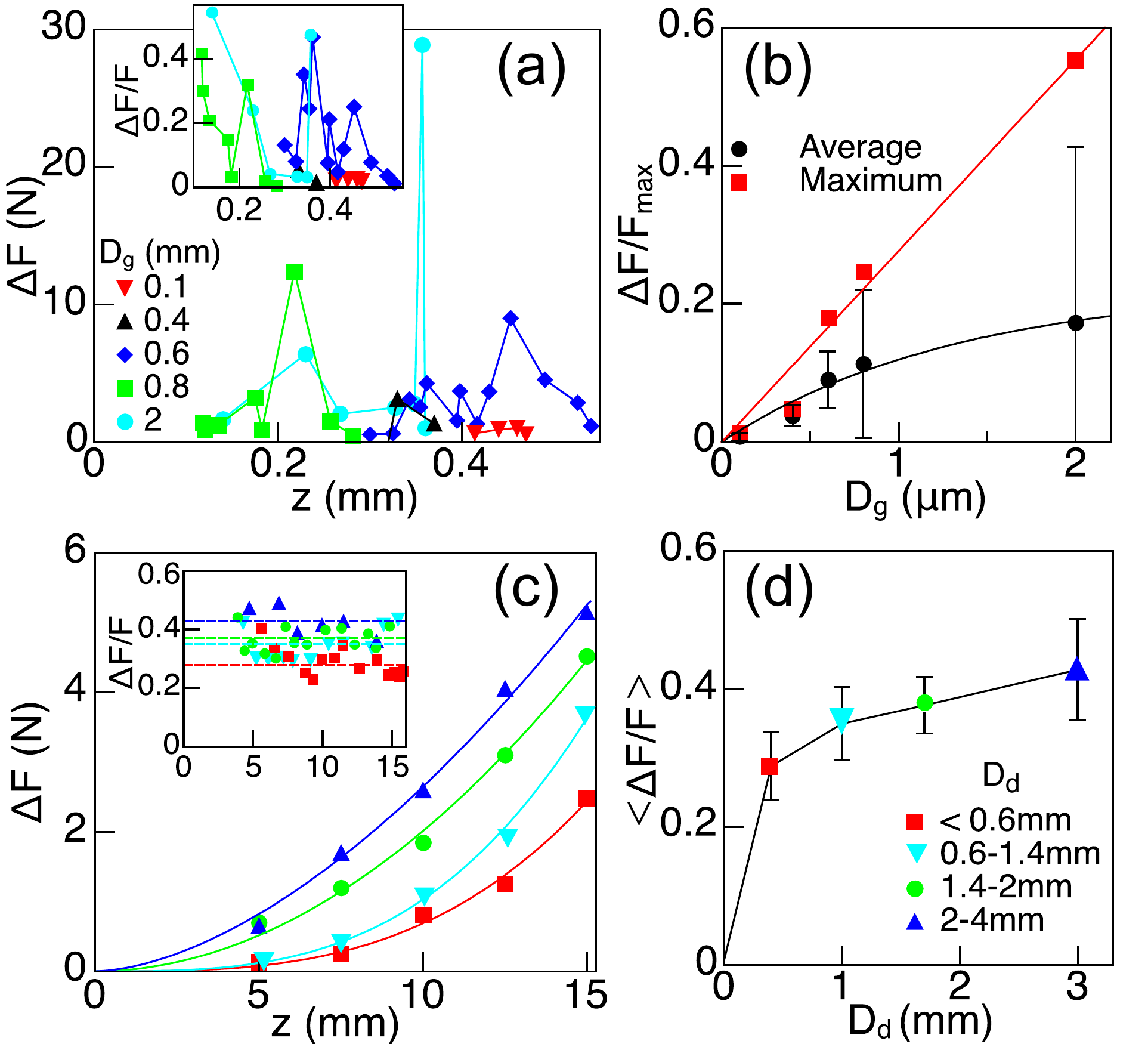}
\vspace*{-0.2cm}
\caption{\textit{(a-b) Glass beads:} a) Amplitude of force drops $\Delta F$ at a given stroke $z$ obtained from Fig. \ref{fig-2}b. Inset: $\Delta F/F$ vs $z$ is erratic.   b) Average and maximum values of $\Delta F/F_{max}$ vs $D_g$, the solid lines correspond to the fitting curves $\Delta F/F_{max}=0.28 D_g$ (red line) and $\Delta F/F_{max}=0.22[1-\exp(-D_g/1.33)]$ (black line). \textit{(c-d) Dust beads:} c) Undulation amplitude $\Delta F$ vs  $z$. Inset: $\Delta F/F$ vs $z$.  Dashed lines indicate the average value, which is plotted as a function of the grain size in (d). $\Delta F$ increases with $z$ and $D_d$.}
\label{fig-5}
\end{figure}
What is considerably modified by the grain size is the magnitude of the abrupt force drops. From the data plotted in fig. \ref{fig-2}b, we measured the force drops $\Delta F$ registered during the compression of glass beads until reaching $F_{max}=100$ N for each grain size, see Fig. \ref{fig-5}a. Note that the magnitude of $\Delta F$ and its relative value $\Delta F/F$ are erratic. Nevertheless, the average and maximum values of the force drops are more marked as the grain size increases (Fig. \ref{fig-5}b), because the abrupt drops are probably related to the particles rearrangement. Since vacancies inside the confined granular column are approximately of the same size than the grains, one needs to apply larger forces to displace larger grains before finding a vacancy.  For the glass beads, the slope of the linear fit in Fig. \ref{fig-5}b (red line) corresponds to 28 kN/m, which indicates that a force of 28 N in average is required to displace a jammed glass bead one millimeter in the confined bed. One can interpret that the small fluctuations in the force are related to local grain sliding events, and the large force drops to the displacement of grains to vacancies inducing the general bed rearrangement. We can discard grain breakage because the applied stress is smaller than the typical glass strength ($\sim 7$ MPa) and its elastic modulus ($\sim 60$ GPa).

For the case of dust particles, the fact that all the curves of $F$ vs $z$ are superimposed in Fig. \ref{fig-4} indicates that the mean value of $F(z)$ is not very sensitive to the grain size. In contrast, $\Delta F$ increases with $z$ following a power-law dependence of the form $\Delta F = A z^p$ (see Fig. \ref{fig-5}c), as the mean value of $F(z)$. The inset indicates that the relative value $\Delta F /F$ remains roughly constant for a given grain size. The mean values $\langle \Delta F /F \rangle$  are represented in this inset by dashed lines and plotted vs $D_d$ in Fig. \ref{fig-5}d. Whereas $\Delta F$ is erratic for glass beads due to stick-slip fluctuations (Fig. \ref{fig-5}a), it increases with $z$ for dust particles (Fig. \ref{fig-5}c). Thus, the absolute and relative values of  $\Delta F$ increase with the grain size for both materials (Figs. \ref{fig-5}b and d).

\subsection{Relaxation process}
\label{sec3-2}
The instantaneous drop of $F(t)$ from $F_{max}$ at $t=0$ in figs. \ref{fig-2}a and \ref{fig-3}c  (color lines) suggests that $F$ decreases exponentially during the relaxation of glass beads. The Maxwell model  has been used to describe the stress relaxation in polymers, polymer matrix composites and soft materials \cite{Papanicolaou2011}.  In a previous work \cite{Katsuragi2020}, we modeled the relaxation of glass beads considering the sum of two exponential terms followed by logarithmic relaxation according to Ref. \cite{Brujic:2005}. Nevertheless, such model would be valid only in a certain range of $t$ considering that the logarithmic dependence diverges when $t \rightarrow \infty$. To avoid such divergence, we fitted the experimental data in Fig. \ref{fig-3}c using the Maxwell model \cite{Papanicolaou2011} with three exponential terms of the form:
$ F/F_{max} = C_1 e^{-t/\tau_1} + C_2 e^{-t/\tau_2} + C_3 e^{-t/\tau_3}$, 
where  $\tau_1$, $\tau_2$ and $\tau_3$ are characteristic (fast, moderate and slow) relaxation times and $C_1$, $C_2$, and $C_3$ are coefficients that in principle satisfy $\sum_{i=1}^{3}C_i=1$. The values of the above parameters obtained from the best fit of the experimental data are indicated in Table 1 as a function of the maximum force load. The fast relaxation time $\tau_1 \sim 20$ s indicates an instantaneous drop of at most $10\%$ of the maximum compression force during the first seconds of relaxation, then, the relaxation is moderate ($\tau_2 \sim300$ s), and finally $F(t)$ decreases slowly to zero during very long relaxation times ($\tau_3 \sim O(10^4$ s). From the values of the coefficients $C_i$, one can notice that $C_3$ saturates  and that the fast relaxation (given by $C_1$ and $C_2$) is less notorious for large values of $F_{max}$. For dust particles, the relaxation process follows basically similar dependence on the grain size with $\tau_3 \rightarrow \infty$.

%

\begin{table}[ht!]
\small
\centering
\label{tab-1}       
\caption{Parameters obtained from the best data fit for glass beads relaxation, for different values of $F_{max}$.}
\vspace*{-0.2cm}
\begin{tabular}{l l l l l l l}
\hline
$F_{max}$& $C_1$ &  $C_2$  &  $C_3$ & $\tau_1(s)$&$\tau_2(s)$ &$\tau_3(s)$ \\\hline
30& 0.034 & 0.045 & 0.91 & 19.7 & 307 & 45900  \\
20 & 0.041 & 0.059 & 0.88 & 23.5 & 302 & 27600 \\
10& 0.062 & 0.071 & 0.84 & 18.4 & 248 & 28500 \\
5 & 0.086 & 0.092 & 0.78 & 19.0 & 292 & 18800  \\
\hline
\end{tabular}
\end{table}

\vspace{-0.7cm}
\section{Conclusions}
\label{sec-4}
By performing compression-relaxation tests on granular columns composed of  glass beads or dust aggregates, we show that the measured force load increases nonlinearly with time, with abrupt drops or periodic undulations, respectively. In both cases, the mean value of $F(z)$ is largely unaffected by the grain size, in the studied range, but the amplitude of the fluctuations increases with the grain diameter. This suggests that the force drops for glass beads are associated to particles rearrangements and stick-slip motion, whereas periodic undulations of the force load for porous dust particles are associated to their deformability and fragility. On the other hand, the Maxwell model with three exponential terms describes reasonably well the relaxation process in both scenarios.  
\\

\noindent \small{Financial support: JSPS KAKENHI Grant No. 18H03679.}

%
%
%

\end{document}